\documentclass[letterpaper, 10 pt, conference]{ieeeconf}

\IEEEoverridecommandlockouts 
\usepackage{graphicx}
\usepackage{epsfig,epstopdf}
\graphicspath{{../pdf/}{../jpeg/}}
\DeclareGraphicsExtensions{.pdf,.jpeg,.png}
\usepackage{amsmath,amssymb,amsfonts,mathrsfs}
\usepackage{multirow}
\usepackage{booktabs}
\usepackage{color}
\usepackage{algpseudocode}
\usepackage{tikz}
\usepackage{color}
\usepackage{cite}
\usepackage{setspace}
\usepackage{algorithm}  
\usepackage{hyperref}
\hypersetup{
colorlinks=true,
linkcolor=black,
filecolor=magenta,      
urlcolor=cyan,
pdftitle={Overleaf Example},
pdfpagemode=FullScreen,
}

\usepackage{float}

\allowdisplaybreaks[4]
\usepackage{theorem}

\newtheorem{assumption}{Assumption}
\newcommand{\R}{\mathbb{R}}
\newcommand{\Z}{\mathbb{Z}}

\newcommand{\bma}[1]{\begin{bmatrix}#1\end{bmatrix}}

\title{\LARGE \bf
Direct Continuous-Time LPV System Identification of Li-ion Batteries via L1-Regularized Least Squares
}

\author{Yang Wang and Riccardo M.G. Ferrari
\thanks{Yang Wang and Riccardo M.G. Ferrari are with the Delft Center for Systems and Control, Delft University of Technology, Delft, 2628CD, Netherlands. Email: {\tt\small \{y.wang-40, r.ferrari\}@tudelft.nl}. This research has been funded by Volvo AB (Sweden) and NWO (Netherlands) under TKI-HTSM grant 21.0056.}%
}

\begin{document}

\maketitle
\thispagestyle{empty}
\pagestyle{empty}

\begin{abstract}
Accurate identification of lithium-ion battery parameters is essential for estimating battery states and managing performance. However, the variation of battery parameters over the state of charge (SOC) and the nonlinear dependence of the open-circuit voltage (OCV) on the SOC complicate the identification process. In this work, we develop a continuous-time LPV system identification approach to identify the SOC-dependent battery parameters and the OCV-SOC mapping. We model parameter variations using cubic B-splines to capture the piecewise nonlinearity of the variations and estimate signal derivatives via state variable filters, facilitating CT-LPV identification. Battery parameters and the OCV-SOC mapping are jointly identified by solving L1-regularized least squares problems. Numerical experiments on a simulated battery and real-life data demonstrate the effectiveness of the developed method in battery identification, presenting improved performance compared to conventional RLS-based methods.
\end{abstract}

\section{INTRODUCTION}
Electric vehicles (EVs) have gained widespread utilization owing to low-carbon policies and sustainable transportation demands \cite{meng2018overview,hannan2017review}. Lithium-ion (Li-ion) batteries are increasingly deployed in EVs for their enhanced energy density, energy efficiency, and extended service life \cite{plett2015battery}. Accurate identification of battery parameters is essential for reliable EV operation, as they are fundamental to estimating battery state of charge (SOC) and state of health (SOH), two critical measures for ensuring EV safety and efficiency \cite{wang2025review}. 


%

Battery parameters are physical quantities that naturally exist in the continuous-time (CT) models derived from physical laws. To retrieve these physical parameters, a CT model is eventually required from the data, even though a discrete-time (DT) model is first obtained. As the concentrations of lithium ions in the battery cell evolve during charging and discharging, the battery parameters exhibit variations over the state of charge (SOC). In addition, the open circuit voltage (OCV) of the battery changes with the SOC with a nonlinear relation \cite{plett2015battery,chen2019novel}, complicating the battery parameter identification task. To address these problems, existing approaches mostly adopt discrete-time (DT) recursive least squares (RLS) based online identification methods \cite{yang2023improved,tian2023lithium}. These approaches construct a DT model of the battery, iteratively update the model parameters using the incoming data, and transform the identified DT model into a CT model to retrieve the battery parameters. Although they track parameter variations, RLS-based methods fail to incorporate functional dependencies of the parameters on the SOC, making the parameters and the SOC misaligned with each other. Moreover, the transformation between the DT model and the CT model adds discretization errors, lowering the identification accuracy \cite{chou1999continuous}.

An alternative approach is to formulate a linear parameter-varying (LPV) model and conduct continuous-time system identification to identify the LPV model parameters. This approach explicitly models battery parameters as functions of the SOC, resolving the misalignment problem in the RLS-based methods. In addition, CT system identification omits the transformation between the DT and the CT model, simplifying the identification process and avoiding discretization errors. Some attempts are made in the literature. \cite{xia2016accurate} developed a local CT-LPV method to identify the SOC-dependent battery parameters, in which the parameters are considered constants at individual SOC points and are identified by applying a linear time-invariant CT method. The local approach is restricted by the number of local models and cannot capture the transient behavior between the working points where the local models are identified. \cite{li2024online} devised a global subspace-based LPV system identification to identify the SOC-dependent battery parameters. However, this method requires the OCV-SOC mapping measured via offline tests, which are inapplicable to installed batteries. \cite{andersson2020continuous} developed a CT-LPV method to identify temperature-dependent battery parameters with the simplified refined instrumental variable method. This approach, nevertheless, only modeled the internal resistance as a variable parameter while treating the dynamic parameters as constants, limiting the model accuracy.


%


In this work, we develop a novel CT-LPV system identification method to jointly identify the SOC-dependent battery parameters and the OCV-SOC mapping. This method models the battery parameters as functions of the SOC and identifies the CT battery parameters directly from the sampled data without relying on a DT model. The dynamic behavior and static OCV are distinctly represented in the input-output (IO) model of the battery derived by latent variable elimination in a latent variable representation. We model the SOC-dependent battery parameters with cubic B-splines to reflect their piecewise nonlinearity and estimate the time-derivatives of signals from sampled data using a state variable filter (SVF) method. Finally, the battery parameters and the OCV-SOC mapping are jointly identified by solving L1-regularized least-squares problems.

\textit{Contributions} The contributions of the paper are summarized as follows:
\begin{itemize}
    \item We develop a CT-LPV system identification method to identify the SOC-dependent battery parameters and the OCV-SOC mapping;
    \item The battery parameters and the OCV curve are modeled with cubic B-splines for enhanced estimation;
    \item The battery parameters and the OCV curve are identified by solving L1-regularized least squares problems;
    \item The proposed method is validated on a simulated battery and real-life data, compared with existing methods.
\end{itemize}

The rest of this paper is structured as follows. Section \ref{sec: model} introduces the battery model used to model Li-ion batteries. Section \ref{sec: method} presents the proposed CT-LPV method for identifying battery parameters and the OCV-SOC mapping. Section \ref{sec: result} demonstrates the results of the proposed method on a simulated battery and real-life data. Section \ref{sec: conclusion} concludes this study.

\section{Model description of Li-ion batteries}\label{sec: model}

\begin{figure}[htbp!]
    \centering
    \includegraphics[width=0.9\linewidth]{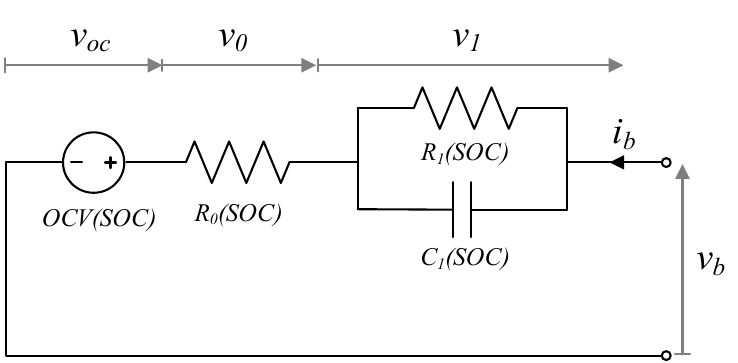}
    \caption{Equivalent circuit model}
    \label{fig:1RC ECM}
\end{figure}

Li-ion batteries are commonly modeled with a first-order equivalent circuit model (ECM), as shown in Figure \ref{fig:1RC ECM}.

This model comprises an internal resistor $R_0$, a resistor-capacitor network $R_1C_1$, and a voltage source $v_{oc}$. $R_0$ models the internal resistance of the battery, $R_1C_1$ represents the battery's polarization process, and $v_{oc}$ emulates the open-circuit voltage (OCV) of the battery. The current $i_b$ and the voltage $v_b$ are the model's input and output, and the charging direction is defined as the positive direction. The continuous-time state-space equations of the ECM can be written as:
\begin{align}
    \dot v_1(t)&=-\frac{1}{\tau_1(z(t))}v_1(t)+\frac{1}{C_1(z(t))}i_b(t)\label{eq: state}\\
    v_b(t)&=v_1(t)+R_0(z)i_b(t)+v_{oc}(z(t)),\label{eq: output}
\end{align}
where $\tau_1(z(t)):=R_1(z(t))C_1(z(t))$ is the time constant of the battery, $z(t)$ is the state of charge (SOC) and $v_1\in\R$ is the polarization voltage across the RC pair. In this model, the battery parameters $R_0,R_1,C_1$ and the OCV $v_{oc}$ vary with the SOC of the battery. The SOC is defined as the amount of charge in the battery in relation to its total capacity. It is computed via coulomb counting as:
\begin{equation}\label{eq:soc}
    z(t)=z(t_0)+\int_{t_0}^t\frac{1}{3600C}i_b(\tau)d\tau,
\end{equation}
where $z(t_0)$ is the initial SOC at time $t_0$, $C$ is the battery capacity in Ampere-hour (Ah), and $3600$ is a factor to convert Ah into Coulombs.

During the identification process, we adopt the following assumptions on the SOC and the intersample behavior of the battery input.
\begin{assumption}[Zero-order-hold input]\label{ass: 1}
    The battery input is generated with a zero-order-hold (ZOH) mechanism.
\end{assumption}
\begin{assumption}[Zero-order-hold SOC]\label{ass: 2}
    The state of charge of the battery is constant during sampling intervals.
\end{assumption}
\begin{assumption}[Constant temperature and age]\label{ass: 3}
    The ambient temperature and the battery age are unchanged during the data collection experiments.
\end{assumption}
Assumption \ref{ass: 1} can be adopted for batteries employed in EVs equipped with digital processors. Assumption \ref{ass: 2} is used for SOC varying slowly and a sufficiently fast sampling rate. It is commonly adopted in the literature \cite{fan2022state,yang2023improved,xia2016accurate}. Assumption \ref{ass: 3} is used to keep simplicity and focus on the essence of the proposed CT-LPV identification method.

The physical parameters of the ECM provide representative values that help us to understand the battery status, and by examining the variation of these parameters, they are critically used to estimate the state of charge and the state of health of the battery.

The aim of this study is to identify the SOC-dependent battery parameters $R_0,R_1,\tau_1$ and the static OCV-SOC mapping $v_{oc}$ from the sampled input and output data $\{i_b,v_b\}$ of the battery.

\section{Continuoust-time linear parameter system identification of Li-ion batteries}\label{sec: method}
To identify the SOC-dependent battery parameters and the OCV-SOC mapping, we first construct an input-output (IO) model of the battery, then utilize cubic B-splines to parameterize model parameters, and finally formulate a regularized least squares problem to identify the battery parameters.
\subsection{Input-output model of the LPV battery system}
The traditional Laplace transform-based approach is inapplicable to derive the IO model of the LPV battery system. To derive an IO model of the system \eqref{eq: state}\eqref{eq: output}, we write the system in a latent variable representation form \cite{toth2010modeling} as:
\begin{align}
    R_L(\xi)v_1(t)=R_W(\xi)w(t)
\end{align}
where $\xi:=\frac{d}{dt}$ is a differential operator, $R_L\in\R^{2\times1}[\xi]$ and $R_W\in\R^{2\times2}[\xi]$ are polynomial matrices:
\begin{align}\notag
    R_L(\xi)=\bma{\xi+\frac{1}{\tau_1(z(t))}\\
    -1},\ R_W(\xi)=\bma{0&\frac{1}{C_1(z(t))}\\
    -1&R_0(z(t))},
\end{align}
and $v_1(t)$ and $w(t):=[v_b(t)-v_{oc}(t),\ i_b(t)]^\top\in\R^2$ are the latent and the manifest variables respectively. 

In this form, we can always find suitable elementary row operations, represented by a unimodular matrix $M(\xi)$ \cite{polderman1997proper}, such that by multiplying $M(\xi)$ by $R_L(\xi),R_W(\xi)$, it holds:
\begin{equation}\label{eq: latent variable elimination}
    M(\xi)[R_L(\xi)\ |\ R_W(\xi)]=\bma{R_L^1&R_W^1\\0&R_W^2},
\end{equation}
where $R_L^1\in\R[\xi]$ and $R_W^1,R_W^2\in\R^{1\times 2}[\xi]$ are transformed polynomial matrices and the (2,1) block matrix is zero.

From the polynomial matrices in \eqref{eq: latent variable elimination}, we can find the manifest behavior, with the latent variable eliminated as:
\begin{equation}\label{eq: manifest behavior}
    R_W^2(\xi)w(t)=0.
\end{equation}
It can be shown that the unimodular matrix
\begin{equation}
M(\xi)=\bma{0&1\\1&\xi+\frac{1}{\tau_1(z(t))}}  
\end{equation}
gives the form \eqref{eq: latent variable elimination} for the battery system. By using \eqref{eq: manifest behavior}, we can write an IO model of the battery system \eqref{eq: state}\eqref{eq: output} as:
\begin{align}\label{eq: IO model}
    \xi&v_b(t)=a_1(z(t))v_b(t)+\xi(b_0(z(t))i_b(t))+\notag\\
    &b_1(z(t))i_b(t)+\xi v_{oc}(z(t))-a_1(z(t))v_{oc}(z(t)),
\end{align}
where $a_1,b_0,b_1$ are the IO model parameters, and they are related to the battery parameters as:
\begin{align}
 a_1(z(t))&=-\frac{1}{\tau_1(z(t))}\\
 b_0(z(t))&=R_0(z(t))\\
 b_1(z(t))&=\frac{R_0(z(t))+R_1(z(t))}{\tau_1(z(t))}.
\end{align}
By identifying the SOC-dependent IO model parameters, we can retrieve the battery parameters using the above relations.


\subsection{Cubic B-splines for model parameterization}
\begin{figure}
    \centering
    \includegraphics[width=1\linewidth]{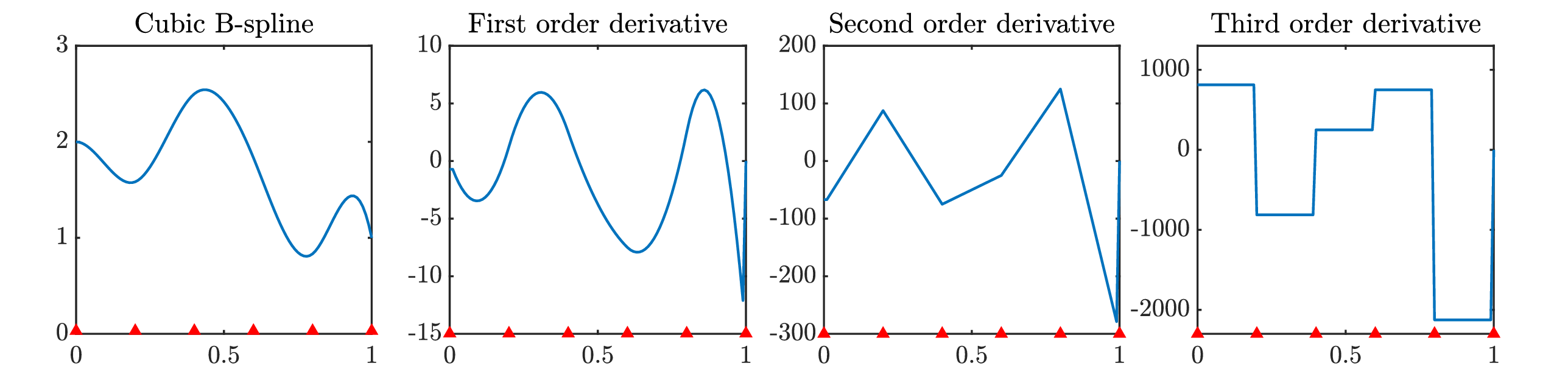}
    \caption{Cubic B-spline and its first to third-order derivatives. Red triangles represent knot positions and mark the derivatives' discontinuities.}
    \label{fig: knot places}
\end{figure}

To identify the SOC-dependent IO model parameters, we parameterize the parameter variations with cubic B-splines. A cubic B-spline is a piecewise polynomial function capable of representing a wide range of functional variations with a low-order model. As noted in the literature, the battery parameters exhibit piecewise nonlinearity over the SOC \cite{shokri2024battery,yang2023improved}, and thus the cubic spline is suitable for modeling these characteristics. The IO model parameters and the OCV term in \eqref{eq: IO model} can be modeled with cubic splines as:
\begin{equation}\label{eq: cubic spline}
    \mathcal{P}(z(t))=\sum_{i=1}^hc_{\mathcal{P}}^ig_i(z(t)),
\end{equation}
where $\mathcal{P}\in\{a_1,b_0,b_1,v_{oc},a_1v_{oc}\}$ are the model parameters, $g_i(z(t))$ is the spline basis evaluated at $z(t)$, $c_{\mathcal{P}}^i$ is the $i$-th control point of the parameter $\mathcal{P}$, and $h$ is the total number of the spline bases. The basis function $g_i$ is defined over a non-decreasing knot vector $Z=[z_0,\ldots,z_{h+3}]$, where $z_0\leq\cdots\leq z_{h+3}$ are knots of the spline. Define $p\in\Z_{\geq 0}$ the degree of the spline. $p=3$ for a cubic spline. The spline basis $g_i$ can be computed recursively via the de Boor-Cox formula \cite{ma2023generalized} as:
\begin{align}\label{eq: boor cox formula}
    g_{i,p}(z(t))=\frac{z(t)-z_i}{z_{i+p}-z_i}g_{i,p-1}(z(t))+\notag\\
    \frac{z_{i+p+1}-z(t)}{z_{i+p+1}-z_{i+1}}g_{i+1,p-1}(z(t)),
\end{align}
where
\begin{equation}
    g_{i,0}(z(t))=\left 
    \{   \hspace{-0.1cm}\begin{array}{ll}
         1&  \text{if}\ z_i\leq z(t) <z_{i+1} \\
         0& \text{otherwise}
    \end{array}
    \right..
\end{equation}
A $p$-th degree spline is $(p-1)$-th order continuous. The $d$-th order derivative of a $p$-degree spline can be computed as:
\begin{align}\label{eq: boor-cox derivative}
    g_{i,p}^{(d)}(z(t))=\frac{p}{z_{i+p}-z_i}g_{i,p-1}^{(d-1)}(z(t))-\notag\\
    \frac{p}{z_{i+p+1}-z_{i+1}}g_{i+1,p-1}^{(d-1)}(z(t)).
\end{align}
The third-order derivative of the cubic spline is a piecewise constant function, with discontinuities occurring at the knot positions, as shown by an example in Figure \ref{fig: knot places}. To mitigate overfitting to measurement noise, in the sequel, we will utilize this feature to regularize the smoothness of splines.

With the model parameters represented with cubic splines \eqref{eq: cubic spline}, we can write the IO model \eqref{eq: IO model} in a regression form as:
\begin{equation}\label{eq: differential equation with spline}
    \xi v_b(t)=\phi^\top(t)c,
\end{equation}
where $\phi\in\R^{5h}$ is a data vector:
\begin{equation}\label{eq:regressors time domain}
    \phi(t)=[(gv_b)(t)\ \xi(gi_b)(t)\ (gi_b)(t)\ \xi g(z(t))\ g(z(t))]^\top,
\end{equation}
$g(z(t))\in\R^h$ is a vector of all spline bases:
\begin{equation}
g(z(t)):=[g_1(z(t)),\ldots,g_h(z(t))],
\end{equation}
$(gv_b),(gi_b)\in\R^h$ in \eqref{eq:regressors time domain} are vectors of $g_iv_b$ and $g_ii_b$ for $i\in\Z_1^h$, and $c\in\R^{5h}$ are the spline control points of the IO model parameters:
\begin{equation}\label{eq: model parameters}
    c=\bma{c_{a_1}^\top&c_{b_0}^\top&c_{b_1}^\top&c_{v_{oc}}^\top&c_{a_1v_{oc}}^\top}^\top,
\end{equation}
in which $c_{\mathcal{P}}\in\R^h$ is a vector of $c_{\mathcal{P}}^i$ in \eqref{eq: cubic spline}.

Since SOC $z(t)$ is a weighted integration of current $i_b$ \eqref{eq:soc}, there can be coefficients $c_1$ and $c_2$ such that $(gi_b)(t)c_1=\xi g(z(t))c_2$ for all $t$. To avoid redundancy in the data vector \eqref{eq:regressors time domain}, we perturb $z(t)$ with a small Gaussian noise, e.g., with a standard deviation of $1\times 10^{-4}$, to improve the numerical stability. This perturbation acts on the scheduling signal and does not affect the identification result. Hereafter, we denote by $\tilde z(t)$ the perturbed SOC.


%

\subsection{State variable filter for time-derivative estimation}
Time-derivatives are essential in CT system identification. A state variable filter (SVF) is a continuous-time low-pass filter enabling prefiltered derivatives of signals from sampled data \cite{garnier2008direct}. A zero-order SVF has an operator form as:
\begin{equation}
    F(\xi)=\frac{1}{(\xi+\nu)^n},
\end{equation}
where $\nu\in\R_{>0}$ is the cut-off frequency and $n\in\Z_{\geq0}$ is the order of the system, $n=1$ in our problem. Denote by $F_j(\xi)=\xi^jF(\xi)$ the $j$-th order SVF, $f_j(t)$ the impulse response of $F_j(\xi)$. Then, the $j$-th order derivative of $v_b$ can be computed as $F_j[v_b](t):=f_j(t)*v_b(t)$, where $*$ is the continuous-time convolution operator. With the SVFs applied to all signals in model \eqref{eq: differential equation with spline}, we can write the IO model with prefiltered signals as:
\begin{equation}\label{eq: filtered IO model}
    F_1[v_b](t)=F[\phi]^\top(t)c
\end{equation}
where $F[\phi](t)\in\R^{5h}$ is a prefiltered data vector:
\begin{align}\label{eq: prefiltered regressors}
    F[\phi](t)=[F_0[gv_b](t)\ F_1&[gi_b](t)\ F_0[gi_b](t)\ \notag\\
    &F_1[g](\tilde z(t))\ F_0[g](\tilde z(t))]^\top,
\end{align}
and $F_0[gv_b]$ is the prefiltered vector of $(gv_b)$ in \eqref{eq:regressors time domain}; other components in \eqref{eq: prefiltered regressors} are defined similarly.


As SVFs are implemented over CT signals and the data $v_b(t),i_b(t),g(z(t))$ are only available as discrete-time samples, the sampled data $\{(gv_b)(t_k),(gi_b)(t_k),g(t_k)\}$ are interpolated into CT signals before conducting the filtering. According to assumptions \ref{ass: 1} and \ref{ass: 2}, we apply ZOH interpolation to sampled data to implement SVFs.

\subsection{Regularized least squares for parameter identification}
Using the prefiltered IO battery model \eqref{eq: filtered IO model}, we can identify the spline control points of the IO model parameters by solving a least squares problem as:
\begin{equation}\label{eq: unregularized LS problem}
    \min_{c}\ \|F_1[V_{b,m}]-F[\Phi_m] c\|_F,
\end{equation}
where $\|\cdot\|_F$ is the Frobenius norm, $F_1[V_{b,m}]\in\R^m$ and $F[\Phi_m]\in\R^{m\times(5h)}$ are the vectors of the prefiltered output $F_1[v_b](t)$ and the data $F_1[\phi](t)$:
\begin{equation}
    F_1[V_{b,m}]=\bma{F_1[v_b](t_1)\\ \vdots \\F_1[v_b](t_m)}, F_1[\Phi_m]=\bma{F_1[\phi]^\top(t_1)\\ \vdots \\F_1[\phi]^\top(t_m)},
\end{equation}
and $F_1[v_b](t_k),F_1[\phi](t_k)$ are the sampled data of the continuous-time signals $F_1[v_b](t),F_1[\phi](t)$ and at the time instant $t_k$, $k\in\Z_1^m$.

To prevent cubic splines from overfitting measurement noise, we regularize the smoothness of the splines in the least squares problem \eqref{eq: unregularized LS problem}. The highest-order derivatives of the splines are piecewise constant functions, as illustrated in Figure \ref{fig: knot places}. The smoothness of the splines increases as the size of the discrepancies between the piecewise segments decreases. This can be achieved by imposing the sparsity on the finite difference of the spline derivatives with an L1-regularization. Specifically, we write an L1-regularized least squares problem for identifying the battery parameters as:
\begin{equation}\label{eq: LS problem regularized}
    \min_{c}\ \|F_1[V_{b,m}]-F[\Phi_m] c\|_F+\sum_{i=1}^3\lambda_i\|\mathcal{D} G_m^{(3)}c_{dyn}\|_1
\end{equation}
where $\lambda_i,i\in\Z_1^3$ are the regularization coefficients, $\|\cdot\|_1$ is the L1-norm, $c_{dyn}\in\{c_{a_1}, c_{b_0}, c_{b_1}\}$ are the spline control points of the dynamic parameters of the IO model \eqref{eq: model parameters}, $\mathcal{D}\in\R^{(m-1)\times m}$ is a finite difference matrix and $ G_m^{(3)}\in\R^{m\times h}$ are the third order derivatives of the cubic spline basis functions:
\begin{equation}
    \mathcal{D}=\bma{1&-1& &\\&\ddots&\ddots&\\& &1 &-1},\  G_m^{(3)}=\bma{g^{(3)}(\tilde z_1)\\g^{(3)}(\tilde z_2)\\\vdots\\g^{(3)}(\tilde z_m)},
\end{equation}
where $g^{(3)}(\tilde z_i)$ is computed by \eqref{eq: boor-cox derivative}, and $\{\tilde z_i\}$ is an ordered sequence of SOC. The L1-regularization $\|\cdot\|_1$ in \eqref{eq: LS problem regularized} imposes most elements of the vectors to be zero.

To ensure the control points of the OCV spline $c_{v_{oc}}$ in the derivative term $\xi v_{oc}$ consistent with that in the bilinear term $a_1v_{oc}$ in the IO model \eqref{eq: IO model}, we utilize the dynamic parameter $\hat c_{dyn}$ identified from \eqref{eq: LS problem regularized} and re-identify $c_{v_{oc}}$ by solving the following regularized least squares problem:
\begin{equation}\label{eq: LS problem ocv}
    \min_{c_{v_{oc}}}\ \|\Psi_m-(F_1[G_m]-F_0[\hat a_1G_m])c_{v_{oc}}\|_F+\lambda_4\|\mathcal{D} G_m^{(3)}c_{v_{oc}}\|_1
\end{equation}
where $\Psi_m=F_1[V_{b,m}]-F[\Phi_m^{(1:3h)}](\hat c_{dyn})$, $F[\Phi_m^{(1:3h))}]\in\R^{m\times(3h)}$ are the first $3h$ columns of $F[\Phi_m]$ in \eqref{eq: unregularized LS problem}, $(\hat c_{dyn})\in\R^{3h}$ are vectors of $\hat c_{dyn}$, and $\hat a_1=G_m\hat c_{a_1}$ is the identified parameter $a_1$. This problem represents all OCV terms with a single set of control points, ensuring consistent $v_{oc}$ throughout the IO model \eqref{eq: IO model}. By solving the regularized least squares problem, we can regularize the smoothness of $v_{oc}$ as well. 

In the next section, we validate the efficacy of the CT-LPV method in identifying battery parameters and the OCV-SOC mapping on a simulated battery and real-life battery data.


\section{Numerical experiments on a simulated battery and real-life data}\label{sec: result}
We first validate our approach on a simulated battery and then apply the method to real-life battery data.

\subsection{Validation on a simulated battery}
\begin{figure}[htbp!]
    \centering
    \includegraphics[width=1\linewidth]{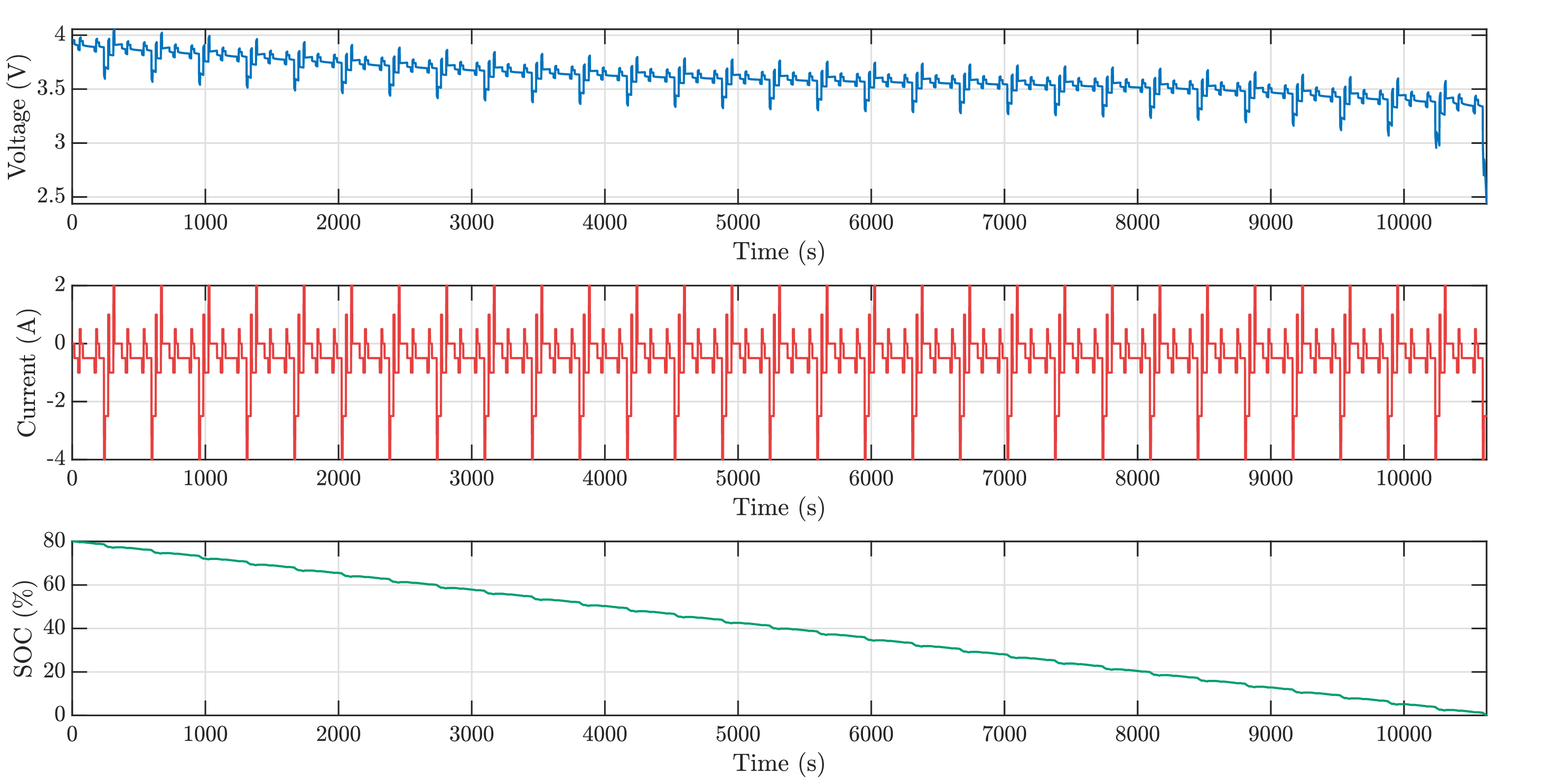}
    \caption{Generated voltage, current, and the computed SOC of the simulated battery under the DST profile.}
    \label{fig:sim_dst_profile}
\end{figure}

To validate the effectiveness of the CT-LPV method in SOC-dependent parameter identification, we generated a simulated battery with the following variable battery parameters:
\begin{align}
    R_0(z)&=0.03\cos(0.3z+2)+0.04/(1+200z^{1.8})+0.1\notag\\
    R_1(z)&=0.3\sin(0.1z+2)+0.6/(1+200z^{1.5})-0.1\notag\\
    \tau_1(z)&=\cos(2z+1)+\sin(5z+1)+18\notag\\
    v_{oc}(z)&=0.03(1.5-z)^{-4}+0.1\log(z+0.01)+3.\notag
\end{align}
These parameters are designed according to real-life battery parameters (Figure 9 in \cite{yang2023improved}) to emulate real battery behavior. To validate the developed method to identify battery parameters and the OCV-SOC mapping from dynamic discharge data, we excited the simulated battery with the Dynamic Stress Test (DST) profile, a current profile that mimics the real-life driving conditions of electric vehicles. The current and voltage of the simulated battery are contaminated with measurement noise, with a standard deviation of $0.01$, estimated from real-life battery measurement data. The generated current, voltage, and the computed SOC of the simulated battery are shown in Figure \ref{fig:sim_dst_profile}.

As the battery is discharged from 80\% to depletion, we selected the number of segments $h-3$ for cubic splines as 80, so that one segment of the spline represents 1\% SOC. We set the cut-off frequency of SVFs as $\nu=1\times 10^{-3}$. The regularized least squares problems \eqref{eq: LS problem regularized} and \eqref{eq: LS problem ocv} are solved in MATLAB with the \texttt{cvx} toolbox \cite{grant2014cvx}. The L1-regularization coefficients $\lambda_i$ in the least squares problems are tuned and set as $\lambda_1=3\times 10^{-5},\lambda_2=5\times 10^{-7},\lambda_3=5\times 10^{-5},\lambda_4=2\times 10^{-5}$. To compare the developed CT-LPV method with an RLS-based method, we applied the discrete-time fixed-memory recursive least squares (FMRLS) \cite{yang2023improved} method to identify the battery parameters. To evaluate the performance of the identification results, we utilize the root mean square error (RMSE) and the Variance-Account-For (VAF) \cite{verhaegen2007filtering} metrics:
\begin{align}
    \text{RMSE}&= \sqrt{\frac{\sum\nolimits_{k=1}^m(y(t_{k})-\hat{y}(t_{k}))^2}{m+1}}\\
    \text{VAF}&=\left(1-\frac{\text{var}(y(t_{k})-\hat y(t_{k}))}{\text{var}(\hat y(t_{k}))}\right)\times 100,
\end{align}
where $y$ is the actual value and $\hat{y}$ is the identified value of the model. A higher VAF indicates a better fit to the data.

\begin{figure}[htbp!]
    \centering
    \includegraphics[width=1\linewidth]{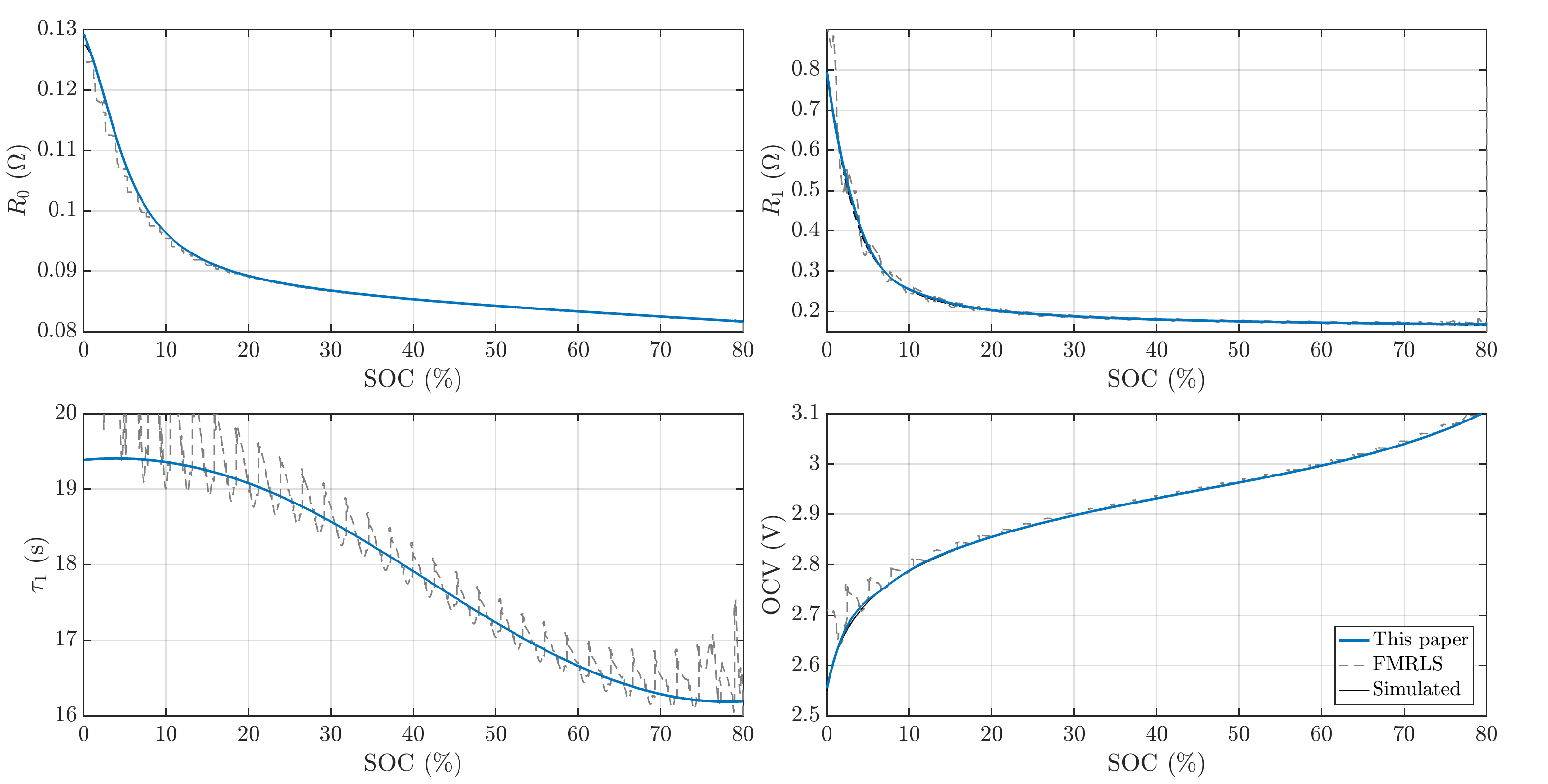}
    \caption{Identified battery parameters of the simulated battery by our CT-LPV method and the FMRLS method.}
    \label{fig:sim_param_voltage_id}
\end{figure}

By applying the CT-LPV method and the FMRLS method to the simulated battery, we obtained the identified parameters shown in Figure \ref{fig:sim_param_voltage_id}. From the figure, we see that the CT-LPV method achieves a smooth curve in the parameters over the SOC, and the identified parameters are consistent with the actual parameters. The FMRLS method contains significant oscillations in the parameters, which are due to the lack of functional dependence on the SOC. On the contrary, the CT-LPV approach enables such a dependence and thus achieves smooth parameters. The RMSEs of the identified parameters by the two methods are tabulated in Table \ref{tab: fitted parameters simdata}. From the table, we see that the CT-LPV approach presented significant improvements in the accuracy of the parameters compared to the FMRLS method. The simulation result indicates that the CT-LPV method is more beneficial than the RLS-based method in identifying the SOC-dependent battery parameters.

\begin{table}[htbp!]
\centering
\tabcolsep 6pt
\caption{RMSE of the identified parameters of the simulated battery}
\begin{tabular}{lcccc}
\hline
Algorithms & $R_0$      & $R_1$ & $\tau_1$   & $v_{oc}$    \\ \hline
CT-LPV       & $5.17\times 10^{-5}$ & 0.0029                 & 0.0025 & 0.0015 \\
FMRLS \cite{yang2023improved}        & 0.0031  & 0.0138                 & 0.7352 & 0.0114 \\ \hline
\end{tabular}
\label{tab: fitted parameters simdata}
\end{table}

With the efficacy of the CT-LPV method validated in simulation, we apply our approach to real-life battery data.

\subsection{Validation on real-life battery data}
\begin{figure}[htbp!]
    \centering
    \includegraphics[width=1\linewidth]{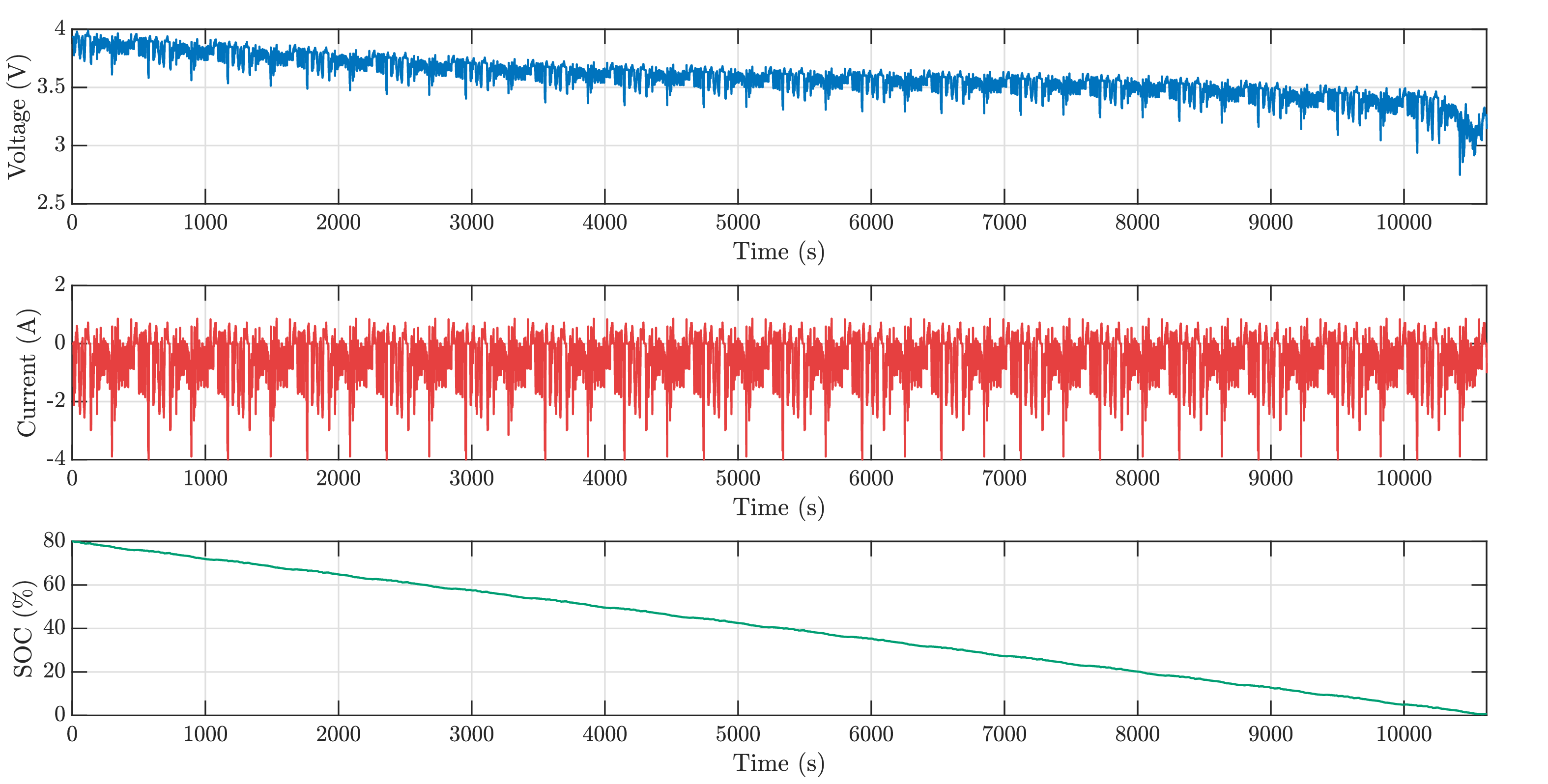}
    \caption{Measured voltage, current, and the computed SOC of the NMC battery under the US06 profile.}
    \label{fig:nmc_us06_profile}
\end{figure}

We adopted the real-life battery data from the CALCE dataset \cite{zheng2016influence}. The test data was collected from an NMC battery, and the battery tests were conducted at a controlled temperature of 25$^\circ$C. The tested battery is discharged with a US06 test profile from 80\% SOC until depletion. The collected voltage, current, and the computed SOC of the tested battery are shown in Figure \ref{fig:nmc_us06_profile}.

\begin{figure}[htbp!]
    \centering
    \includegraphics[width=1\linewidth]{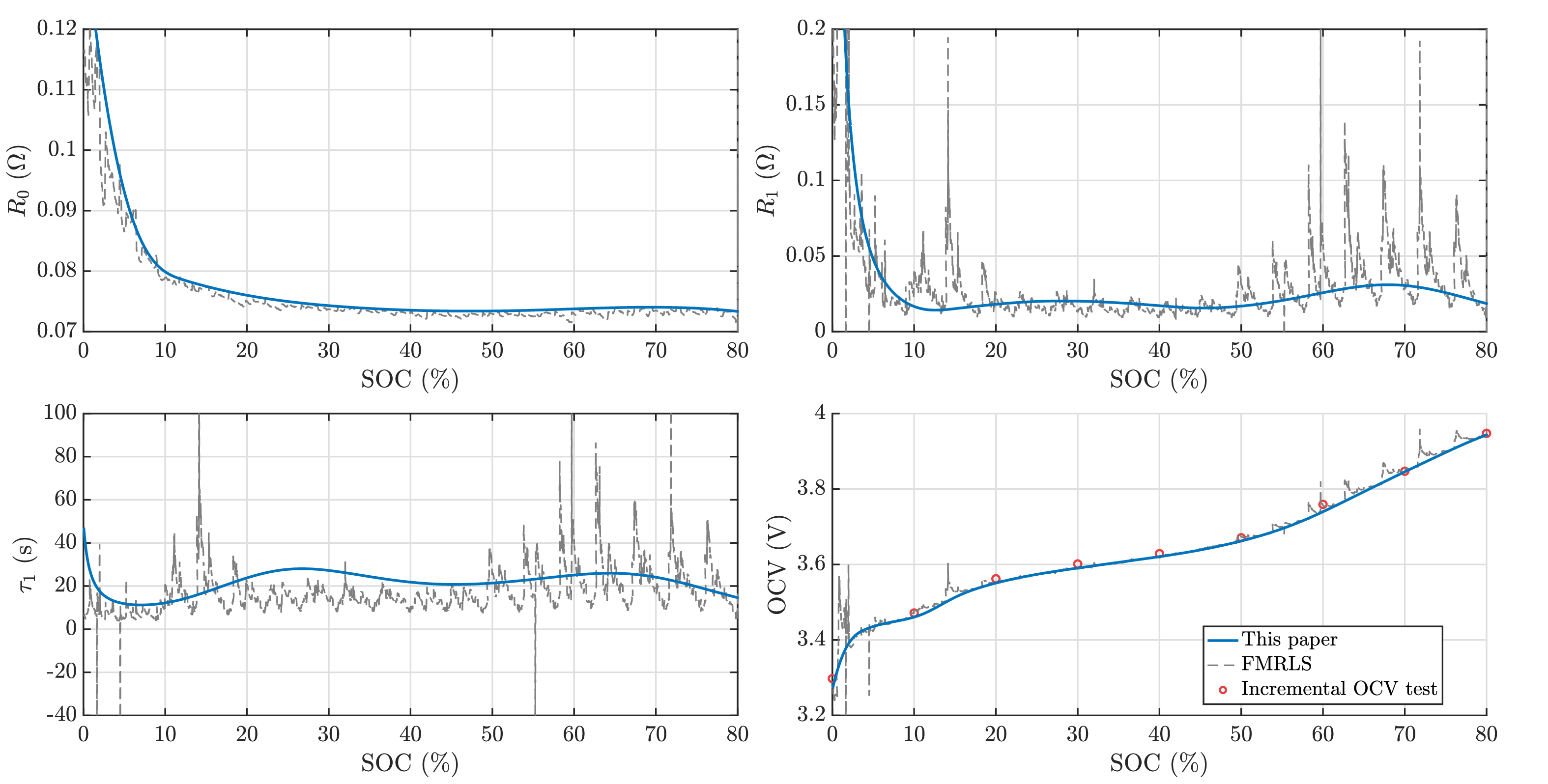}
    \caption{Identified battery parameters of the NMC battery by our CT-LPV method and the FMRLS method.}
    \label{fig:nmc_us06_param_id}
\end{figure}

We applied the CT-LPV method and the FMRLS method to the collected dataset. Since the battery is discharged from 80\% to depletion, we selected the number of segments $h-3$ for cubic splines as 80, so that one segment of the spline represents 1\% SOC. The cut-off frequency of the SVFs was $1\times 10^{-4}$. The battery parameters identified by two methods are shown in Figure \ref{fig:nmc_us06_param_id}. We can see from the figure that our approach presents smoother and less oscillatory estimates of the battery parameters compared to the FMRLS method. The CT-LPV method also effectively tracks the OCV curve as validated by the OCV values obtained from offline OCV tests. This result verifies the consistency of the OCV-SOC mapping identification.

\begin{figure}[htbp!]
    \centering
    \includegraphics[width=1\linewidth]{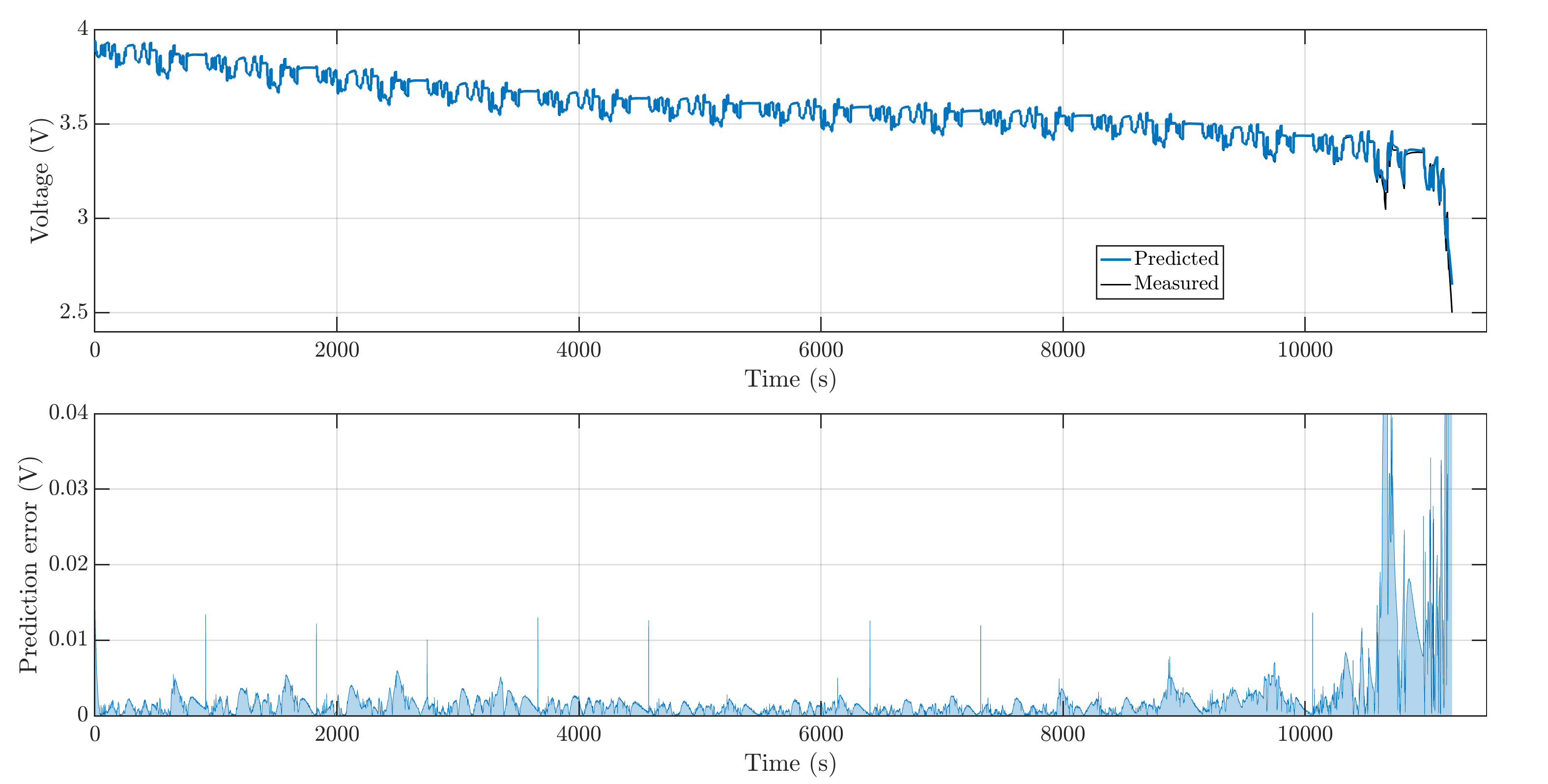}
    \caption{Terminal voltage prediction of the battery model identified by CT-LPV method on the BJDST test dataset.}
    \label{fig:nmc_bjdst_prediction}
\end{figure}

With the battery parameters identified by the CT-LPV method, we examine the predictability of the identified battery model on a new test dataset. We predicted the battery voltage under a BJDST test profile and compared the predicted value with the actual measurements taken from the battery. The predicted terminal voltage and the prediction error of the model on the test dataset are shown in Figure \ref{fig:nmc_bjdst_prediction}. The RMSE of the voltage prediction is 8.5039 mV, and the VAF of the prediction is 99.74\%, improved over that of the FMRLS method, as shown in Table \ref{tab:rmse_vaf_nmc_bjdst_prediction}. The accurate prediction of the terminal voltage on a test dataset validates the battery parameters identified by our CT-LPV method.

\begin{table}[htbp!]
\caption{RMSEs of the terminal voltage prediction of the identified battery models on the BJDST dataset}
\centering
\tabcolsep 17pt
\begin{tabular}{lcc}
\hline
Algorithms & RMSE (mV) & \multicolumn{1}{c}{VAF (\%)} \\ \hline
CT-LPV       & 8.5039    & 99.74    \\
FMRLS \cite{yang2023improved}       & 31.6888    & 96.63    \\\hline                 
\end{tabular}
\label{tab:rmse_vaf_nmc_bjdst_prediction}
\end{table}

\section{Conclusions}\label{sec: conclusion}
In this paper, we develop a continuous-time LPV system identification method to identify the SOC-dependent battery parameters and the OCV-SOC mapping of Li-ion batteries. The CT-LPV model is parameterized with cubic B-splines, and the time-derivatives of signals are estimated with the state variable filter. The battery parameters and the OCV function are jointly identified by solving L1-regularized least squares problems. Validation on a simulated battery and real-life battery data demonstrates that the CT-LPV method achieves a better performance than the RLS-based method in terms of parameter smoothness and accuracy. The battery parameters identified by the CT-LPV approach present an accurate prediction of battery voltage on a test dataset. In future work, we will leverage the identified battery parameters to estimate the SOC and SOH based on dynamic discharge data.



\bibliographystyle{ieeetr}
\bibliography{library}






\end{document}